\documentclass{article}

\newtheorem{Theorem}{Theorem}
\usepackage{epsfig}
\usepackage{oldgerm}

\begin{document}

\markboth{Jose L. Martinez-Morales}
{The Einstein's linear equation of a space-time with a homogeneous section of low dimension}
\title{The Einstein's linear equation of a space-time with a homogeneous section of low dimension}

\author{Jose L. Martinez-Morales}

\maketitle

\noindent{Institute of Mathematics\\
Autonomous National University of Mexico\\
A.P. 273, Admin. De corers \#3\\
Cuernavaca, Morels 62251\\
Mexico\\
martinez@matcuer.unam.mx}
\begin{abstract}
The Einstein's linear equation of a small perturbation in a space-time with a homogeneous section of low dimension, is studied. For every harmonic mode of the horizon, there are two solutions which behave differently at large distance $r$. In the basic mode, the behavior of one of the solutions is ${{(-{r^2}+{t^2})}^{\frac{1-n}{2}}}$ where $n$ is dimension of space. These solutions occur in an integral form.

In addition, a main statement of the article is that a field in a black hole decays at infinity according to a universal law. An example of such a field is an eigentensor of the Einstein's linear operator that corresponds to an eigenvalue different from Zero. Possible applications to the stability of black holes of high dimension are discussed.

The analysis we present is of a small perturbation of space-time. The perturbation analysis of higher order will appear in a sequel. We determine perturbations of space-time in dimension 1+$n\ge$ 4 where the system of equations is simplified to the Einstein's linear equation, a tensor differential equation. The solutions are some integral transformations which in some cases reduce to explicit functions.

We perform some perturbation analysis and we show that there exists no perturbation  regular everywhere outside the event horizon which is well behaved at the spatial infinity. This confirms the uniqueness of vacuum space-time within the perturbation theory framework.

Our strategy for treating the stability problem is applicable to other space-times of high dimension with a cosmological constant different from Zero.
\end{abstract}


{PA CS numbers: 04.70.Dy}
\section{Introduction}
This paper is a discussion concerning gravitational perturbation theory. We intend to investigate a linearised perturbation of metrics of the static black hole space-time  in arbitrary dimension, and discuss the important case of  Wheeler  equations. Much attention is paid to the asymptotic behaviour at the spatial infinity. For a linearised perturbation, at least in 4d the most interesting case is the uniform boundness of the wave function in time defined by the integral curves of the time-like Killing field. The Price law is the prime concern in this context for black hole physics. General analysis of stability of static black holes in $D$ dimensions has been performed by various authors, in particular   \cite{KodamaStability}, and in this respect we present new results.

The motivation and the scientific advance made in this work are stated in the theorems \ref{Theorem 1} and \ref{Theorem 2}.

Why a fluctuation of a black hole is regular? This is an important problem in gravitation which has been studied for a long time. Although we do not have yet a definite answer to this question we have an answer to the question of a stable fluctuation of a black hole which is summarized as the instability of gravitation.

In local particle physics, the assumption of the asymptotic flatness is a good approximation. However, the success of the inflationary universe model implies that the cosmological constant can not be neglected in the early universe even locally. Further, the recent unifying theories suggest the possibility that the universe has more than four dimensions on microscopic scales \cite{Randall, Antoniadis}. Furthermore, mini black holes might be produced in elementary particle processes in colliders as well as in an event of a cosmic shower of high energy. Therefore, it is a quite important problem determining whether a fluctuation of a black hole is regular with a cosmological constant different from Zero and in high dimensions.

We want to investigate the stability of space-time in the framework of a gauge invariant formalism for gravitational perturbations. This formalism has been developed in \cite{Ishibashi}. Perturbations are classified into three s: tensor, vector and scalar, according to their behavior on the low dimensional section of the space-time. The vector and scalar  modes that correspond, respectively, to axial and polar modes in the four dimensional case.

The stability of solutions of a nonlinear system is a nonlinear concept and it is difficult to prove \cite{20a}. Therefore we are only going to consider the sign of eigenvalues of the linear partial differential equations.
\subsection{Statement}
We now discuss what kind of new information is obtained from the perturbation theory. We begin by discussing properties of the eigentensors of the Laplace's operator for a space-time with possible applications to the stability of black holes of high dimension. As far as we know the question of the stability of black holes of any dimension, generalizing the original work of  Wheeler, has been analyzed in detail in \cite{KodamaStability}, and also by Gibbons and collaborators, and their work leaves little room for new developments. The analog of Wheeler  equations have been  recently derived for a black hole Sitter space-time of D dimensions and they are summarized by a master equation (see for example \cite{Kodama, Ishibashi, Konoplya}). 
 
The purpose of the paper and the new results are exposed in Section \ref{Section 4}. As a main statement of the article we find that any eigentensor of the Einstein's linear operator in a black hole that corresponds to an eigenvalue different from Zero decays at infinity in a universal way.

Consider
\begin{itemize}
\item two derivable functions of the variable $r$, denoted by $g_ {00}$ and $g_ {11}$,
\item a compact Riemann manifold ($M\sp{n-1}$, {\textgoth h}) of dimension $n$-1 and
\item the metric
\begin{equation}
\label{1}
\hbox{\textgoth g}=-g_{00}d t\otimes d t+g_{11}d r\otimes d r+r\sp2\hbox{\textgoth h},
\end{equation}
defined on the Cartesian product of two lines and $M\sp{n-1}$.
\end{itemize}
\begin{Theorem}
\label{Theorem 1}
For every harmonic mode of the horizon, there are two solutions of the Einstein's linear equation that behave differently at large. In the basic mode, the behavior of one solution is ${{(-{r^2}+{t^2})}^{\frac{1-n}{2}}}$.
These solutions occur in integral form. Moreover, dependence on $r$ of an eigentensor of the Einstein's linear operator that corresponds to an eigenvalue different from Zero is regular at infinity.
\end{Theorem}
In the next section we give a general explanation of our formalism for gravitational perturbation and the basic strategy of our stability analysis. At this stage our argument is not limited to the case of black holes, it is equally applicable to any case of space-time with a cosmological constant different from Zero. In section \ref{Section 4} we solve the Einstein's linear equation of a space-time with a homogeneous section of low dimension.
\section{Solution of the Einstein's linear equation}
\label{Section 4}
This section is devoted to constructing an approximate solution of the Einstein's linear equation of a space-time with a homogeneous section of low dimension (\ref{13}).

Consider
\begin{itemize}
\item a function $f$ of the variable $r$ and
\item an operator $O$ on a vector space (for example, the Laplace operator acting on functions on the sphere).
\end{itemize}
Let us consider a field with configuration manifold $M\sp{1+n}$ with global coordinates $r$, $t$, $\chi$ described by a singular Euler-Lagrange equation,
\begin{equation}
\label{13}
\left(\frac{1}{{g_{00}}(r)}\frac{\partial\sp2}{\partial t\sp2}-\frac{{r^{1-n}}}{{\sqrt{{g_{11}}(r)}} {\sqrt{{g_{00}}(r)}}}\frac\partial{\partial r}\left(\frac{{r^{n-1}} {\sqrt{{g_{00}}(r)}}}{{\sqrt{{g_{11}}(r)}}}\frac\partial{\partial r}\right)+\Big(\frac{O}{{r^2}}+f(r)\Big)\right)\phi=0.
\end{equation}
In absence of explicit solutions of the Einstein's linear equation, the best we can do is to construct a transformation. On the other hand, such transformation must have the remarkable property that in special regimes, namely $\omega(r, t, \chi ) \approx k$, the variables $(r, t, \chi)$ and $(k, \omega, e)$ form a basis of observable variables outside the shell of the gravitational field even if the solution of the Einstein's linear equation is not known.

The fundamental part of equation (\ref{13}), relevant for perturbation theory, are the two first terms.

Suppose that the operator $O$ has a complete set of eigenvectors $\{E_e\}$.

We denote by $B_o$ the Bessel function of order $o$. If $o$ is a negative integer number then $B_o$ denotes the Bessel function of the second type and order $o$.

If $\{E_e\}$ is a discreet set (continuous set) then $\sum_e$ denotes a sum (an integral).

Set
\[
o=\pm{\sqrt{e+{\left(-1+\frac n2\right)}^2}}.
\]
We propose a solution of equation (\ref{13}) of the form
\begin{equation}
\label{14}
{r^{1-\frac{n}{2}}}\sum_e E_e\int_{-\infty}\sp\infty\exp\Big({\sqrt{-1}}t \omega \Big)\int_0\sp\infty {B_o}(k r) F(e,k,\omega )d k d\omega.
\end{equation}
A vector $E_e$ satisfies
\begin{equation}
\label{15}
OE_e=e E_e.
\end{equation}
The Bessel's functions satisfy with
\begin{eqnarray}
\label{16}
2B_o'&=&B_{o-1}-B_{1+o},\\
\label{17}
\frac{2o}x B_o(x)&=&B_{o-1}(x)+B_{1+o}(x).
\end{eqnarray}
We substitute (\ref {14}) into (\ref {13}).

We use (\ref {15}), (\ref {16}) and (\ref {17}).

(\ref {14}) becomes
\[
{r^{-n/2}}\sum_e E_e\int_{-\infty}\sp\infty\exp\Big({\sqrt{-1}}t \omega \Big)\int_0\sp\infty F(e,k,\omega ),
\]
\[
\Bigg(\frac{k r B_o'(x)|_{x=k r} \frac d{d r} {{\log\big(\frac{{g_{11}}(r)}{{g_{00}}(r)}\big)}^{\frac{1}{2}}}}{{g_{11}}(r)}+{B_o}(k r) \Bigg(\frac{e(1-\frac{1}{{g_{11}}(r)})}{r}+r \bigg(f(r)+\frac{{k^2}}{{g_{11}}(r)}
\]
\begin{equation}
\label{18}
-\frac{{{\omega}^2}}{{g_{00}}(r)}\bigg)+\frac{\frac d{d r} {{\log\big(\frac{{g_{00}}(r)}{{g_{11}}(r)}\big)}^{\frac{1}{2}(-1+\frac{n}{2})}}}{{g_{11}}(r)}\Bigg)\Bigg)=0.
\end{equation}
We suppose that for every $e$ and every $\omega$:
\[
\lim_{k\to0}k F(e,k,\omega ) B_o'(x)|_{x=k r}=\lim_{k\to\infty}k F(e,k,\omega ) B_o'(x)|_{x=k r}.
\]
We integrate by parts $\int_0 \sp \infty k F (e, k, \omega) B_o'(x)|_ {x=k r} d k$.

(\ref {13}) becomes
\[
{r^{-n/2}}\sum_e E_e\int_{-\infty}\sp\infty\exp\Big({\sqrt{-1}}t \omega \Big)\int_0\sp\infty {B_o}(k r),
\]
\[
\Bigg(F(e,k,\omega ) \Bigg(\frac{e(1-\frac{1}{{g_{11}}(r)})}{r}+r \bigg(f(r)+\frac{{k^2}}{{g_{11}}(r)}-\frac{{{\omega}^2}}{{g_{00}}(r)}\bigg)+\frac{\frac d{d r} {{\log\big(\frac{{g_{11}}(r)}{{g_{00}}(r)}\big)}^{\frac{n}{4}}}}{{g_{11}}(r)}\Bigg),
\]
\[
+\frac{k \frac d{d r} {{\log\big(\frac{{g_{11}}(r)}{{g_{00}}(r)}\big)}^{\frac{1}{2}}} \frac\partial{\partial k}F(e,k,\omega )}{{g_{11}}(r)}\Bigg)=0.
\]
Therefore:
\[
\Bigg(F(e,k,\omega ) \Bigg(\frac{e(1-\frac{1}{{g_{11}}(r)})}{r}+r \bigg(f(r)+\frac{{k^2}}{{g_{11}}(r)}-\frac{{{\omega}^2}}{{g_{00}}(r)}\bigg)+\frac{\frac d{d r} {{\log\big(\frac{{g_{11}}(r)}{{g_{00}}(r)}\big)}^{\frac{n}{4}}}}{{g_{11}}(r)}\Bigg),
\]
\begin{equation}
\label{19}
+\frac{k \frac d{d r} {{\log\big(\frac{{g_{11}}(r)}{{g_{00}}(r)}\big)}^{\frac{1}{2}}} \frac\partial{\partial k}F(e,k,\omega )}{{g_{11}}(r)}\Bigg)=0.
\end{equation}
We define
\begin{equation}
\label{20}
\omega(k)=\surd \Bigg(\Bigg(f(r)+\frac{e(1-\frac{1}{{g_{11}}(r)})}{{r^2}}+\frac{{k^2}}{{g_{11}}(r)}\Bigg) {g_{00}}(r)+\frac{(\frac n2+o) \frac d{d r} \frac{{g_{00}}(r)}{{g_{11}}(r)}}{2 r}\Bigg).
\end{equation}
Then Equation \ref {19} is as
\begin{equation}
\label{21}
\Bigg(F(e,k,\omega ) \Bigg(\frac{{{g_{11}}(r)}({\omega }^2-\omega(k)^2)}{{r\frac d{d r} {{\log\sqrt{\frac{{g_{11}}(r)}{{g_{00}}(r)}}}}}}+o\Bigg)-k\frac\partial{\partial k}F(e,k,\omega )\Bigg)=0.
\end{equation}
We have a definition of $\omega(k)$ in equation (\ref{20}), the right hand side clearly depends on $r$. However, this function is used in (\ref{21}) to construct an expression for $F(e,k,\omega)$ which we claim to be independent of $r$. Assume the support of the function $ F $($ e $, $ \omega $, $ k $) lies in a narrow neighborhood of the curve
\[
\omega^2=\omega(k)^2.
\]
This is crucial for the rest of the derivation. Then Equation \ref {21} is approximately equal to
\[
o F(e, \omega, k)-k \frac\partial{\partial k}F(e, \omega, k)=0.
\]
Therefore, there exists a function $ G $ such that
\[
F(e, \omega, k)={k^o} G(e, \omega).
\]
Therefore, (\ref {14}) is as
\begin{equation}
\label{22}
{r^{1-\frac{n}{2}}}\sum_e E_e\int_{-\infty}\sp\infty\exp\Big({\sqrt{-1}}t \omega \Big)G(e, \omega)\int_0\sp\infty {k^o}{B_o}(r k)  d k d\omega.
\end{equation}
Set
\[
t=0.
\]
(\ref {18}) becomes
\begin{equation}
\label{23}
{r^{1-\frac{n}{2}}}\sum_e E_e\int_{-\infty}\sp\infty G(e, \omega)\int_0\sp\infty {k^o}{B_o}(r k)  d k d\omega.
\end{equation}
Therefore (\ref {22}) solves (\ref {13}) with initial data (\ref {23}).

Notice that  Bessel functions are unusual in the standard formalism. In the 4d case, the perturbation equation after separation of variables becomes ordinary differential equations with two regular singularities and one Irregular singularity at infinity while Bessel has only one regular singularity and a irregular one.
\subsection{Decay of the field at infinity}
Suppose that
\[
\lim_{r\to\infty}f(r)=0\quad\hbox{ and }\quad\lim_{r\to\infty}g_{**}=1,\quad*=0, 1.
\]
Then
\[
\lim_{r\to\infty}\omega(k)=k.
\]
We calculate
\[
\int_0\sp\infty\exp\Big({\sqrt{-1}}t k \Big) {k^o}{B_o}(k r) d k=-\frac{{{(-2 r)}^o} {{(-{r^2}+{t^2})}^{-\frac{1}{2}-o}} \Gamma\big(\frac{1}{2}+o\big)}{{\sqrt{-\pi}}}.
\]
Consider a number $l$.

Set
\[
e=l (-2+l+n).
\]
Then we have proved:
\begin{Theorem}
\label{Theorem 2}
The field at infinity (\ref {18}) of a black hole decays like
\[
{r^l} {{\big(-{r^2}+{t^2}\big)}^{\frac{1-n}{2}-l}}.
\]
\end{Theorem}
This result improves the cited works.

The previous discussion applies to a class of  space-times which are flat at the spatial infinity \cite{20a}. In them we have the following properties \cite{19a, 19b, 19c, 8a}:

1) The imposition of suitable boundary conditions on the field reduces the symmetries at the spatial infinity to the Poincare group. The asymptotic implementation of the Poincare group makes possible the general relativistic definition of angular momentum and the matching of general relativity with particle physics.

2) These boundary conditions require that a foliation is associated with an acceptable splitting of space-time. Such foliation must tend to the spatial hyperplanes orthogonal to the momentum in a way which is independent of the direction. This property is concretely enforced by using a technique for the selection of space-times which admit flat coordinates at the spatial infinity \cite{16a, 16b}.

These facts entail that there is an effective evolution in the mathematical time which parametrizes the foliation with any splitting of space-time.
\subsection{Summary and discussion}
Our space-time is a general one with a homogeneous section of low dimension, and $\hbox{\textgoth h}$ is the metric of a general Einstein manifold with Rici curvature
\[
{\hbox{\rm Rici}_{\hbox{\textgoth h}}}_{i j}=\frac{{S}}{n-1}h_{i j}.
\]
As a reason to concentrate on the analysis of the tensor case we argue that a general black hole needs to be stable with respect to vector perturbations and scalar perturbations, due to the fact that the higher dimensional standard black hole is believed to be stable with respect to these perturbations, at least in the case when the event horizon takes the form of a compact Einstein manifold of positive curvature. Our argument is based on two points. Firstly, due to the fact that the curvature tensor does not appear in the equations of the scalar perturbations and the vector perturbations, the stability with respect to these modes is determined only by the spectrum of the Einstein manifold. Second, the smallest eigenvalues of the Laplace's operator for these modes is always greater or equal to those for the horizon. Hence, the standard black hole is the most unstable. Due to the fact that the zero mode does not affect the stability analysis we consider as the smallest eigenvalue for the scalar mode, the second smallest one. Furthermore, the first point implies that even for a general space-time, the equations are valid if we replace the eigenvalues $l(l+n-2)$ of $S\sp{n-1}$ by those of the Einstein manifold. Thus, our results in this paper confirm previous analysis on a firmer basis, giving a proof for their argument summarized above.

Regarding the stability of a general space-time with a homogeneous section of low dimension with respect to the tensor case, the perturbation gets coupled to the curvature. However, the master equation is still valid if we replace $l(l+n-2)-2$ by $\lambda_L$-2$\frac{{S}}{n-2}$, where $\lambda_L$ is the eigenvalue of the Li operator of the Einstein manifold, as has been shown by Gibbons \cite{GibbonsHartnoll}. Hence, we obtain a sufficient condition for stability:
\[
\lambda_L \ge (3n-5)\frac{{S}}{(n-2) (n-1)}.
\]
This condition coincides with the condition obtained by Gibbons.

Furthermore, we have proved that for every harmonic mode of the horizon, there are two solutions of the Einstein's linear equation that behave differently at large. In the basic mode, the behavior of one solution is ${{(-{r^2}+{t^2})}^{\frac{1-n}{2}}}$. These solutions occur in integral form. Moreover, the dependence on $r$ of an eigentensor of the Li operator that corresponds to an eigenvalue different from Zero is regular at infinity. On the other hand, the field at infinity of a black hole decays like ${r^l} {{\big(-{r^2}+{t^2}\big)}^{\frac{1-n}{2}-l}}$.

\end{document}